\documentclass[12pt,preprint]{aastex}

\slugcomment{Accepted to ApJ (April, 3 2012)}

\shorttitle{The youngest outflow discovered in MMS 6/OMC-3}
\shortauthors{Takahashi et al.}

\begin{document}

\title{Spatially Resolving Substructures within the Massive Envelope around an Intermediate-mass Protostar: MMS 6/OMC-3}

 \author{Satoko Takahashi,\altaffilmark{1} Kazuya Saigo,\altaffilmark{2} Paul T.P. Ho\altaffilmark{1,3} and Kengo Tomida,\altaffilmark{2,4}}

 \altaffiltext{1}{Institute of Astronomy and Astrophysics, Academia Sinica, P.O. Box 23-141, Taipei 106, Taiwan; satoko\_t@asiaa.sinica.edu.tw;}
 \altaffiltext{2}{National Astronomical Observatory of Japan, Osawa 2-21-1, Mitaka, Tokyo 181-8588, Japan;}
 \altaffiltext{3}{Harvard-Smithsonian Center for Astrophysics, 60 Garden Street Cambridge, MA 02138, U.S.A.}
 \altaffiltext{4}{Department of Astronomical Science, Graduate University for Advanced Studies, national Astronomical Observatory of Japan, Osawa 2-21-1, Mitaka, Tokyo 181-8588, Japan;}

\begin{abstract}
With the Submillimeter Array
\footnote{The Submillimeter Array is a joint project between the Smithsonian Astrophysical Observatory and the Academia Sinica Institute of Astronomy and
 Astrophysics and is funded by the Smithsonian Institution and the Academia Sinica.}
 (SMA), the brightest (sub)millimeter continuum source in the \hbox{OMC-2/3} region,
\hbox{MMS 6}, has been observed in the \hbox{850 $\mu$m}
continuum emission with approximately 10 times better angular resolution than previous studies (\hbox{${\approx}0.3''$}; \hbox{${\approx}120$} AU at Orion).
The deconvolved size, the mass, and the column density of \hbox{MMS 6-main} are estimated to be \hbox{0$''$.32$\times$0$''$.29} \hbox{(132 AU$\times$120 AU)},
\hbox{0.29 M$_{\odot}$}, and \hbox{2.1$\times$10$^{25}$ cm$^{-2}$}, respectively.
The estimated extremely high mean number density, \hbox{1.5$\times$10$^{10}$ cm$^{-3}$}, suggests that \hbox{MMS 6-main}
is likely optically thick at \hbox{850 $\mu$m}.
We compare our observational data with three theoretical core models:
prestellar core, protostellar core + disk-like structure, and first adiabatic core.
These comparisons clearly show that the observational data cannot be modeled as a simple prestellar core with a gas temperature of \hbox{20 K}.
A self-luminous source is necessary to explain the observed flux density in the (sub)millimeter wavelengths. Our recent detection of a very compact and energetic outflow in the CO (3--2) and HCN (4--3) lines, supports the presence of a protostar.
We suggest that \hbox{MMS 6} is one of the first cases of an intermediate mass protostellar core at an extremely young stage.
In addition to the \hbox{MMS 6-main} peak, we have also spatially resolved a number of spiky structures and sub-clumps, distributed over the central 1000 AU.
The masses of these sub-clumps are estimated to be 0.066--0.073 M$_{\odot}$, which are on the order of brown dwarf masses.
Higher angular resolution and higher sensitivity observations with ALMA and EVLA will reveal the origin and nature of these structures such as whether they are originated from fragmentations, spiral arms, or inhomogeneity within the disk-like structures/envelope

\end{abstract}

\section{INTRODUCTION}
The study of the earliest phase of star formation, especially around the prestellar core and Class 0 phases, is essential to understand star formation.       This period includes the important phases such as the prestellar core phase, first adiabatic core phase, and the subsequent formation of the protostar (Larson 1969; Masunaga \& Inutsuka 2000, Bate 1998; 2010; Machida et al. 2006; Saigo et al. 2008; Schoneke \& Tscharnuter 2011). Studying prestellar cores would provide clues for understanding the initial conditions of star formation.
First adiabatic cores are able to launch outflows (Tomisaka 1998; 2002; Machida et al. 2006; Tomida et al. 2010a), and may be related to the formation of binary systems (Bonnell 1994; Matsumoto et al. 2003; Commercon et al. 2010). In the protostellar phase (i.e., after 2nd collapse), substantial mass accretes onto
the central star. Hence this phase would be a key to determine the final stellar mass. However, candidates of this youngest stage of protostars are still very limited in number, and theoretical models are not well constrained as yet.

The Orion Molecular Cloud-2/3 is located at the northern part of the Orion giant molecular cloud A ($d{\approx}414$ pc; e.g., Menten et al. 2007), and is one of the nearest active star forming regions (OMC-2/3; e.g., Aso et al. 2000; Chini et al. 1997; Lis et al. 1998; Johnstone \& Bally 1999; Nielbock et al. 2003;
Takahashi et al. 2006; 2008a,b 2009; Peterson \& Megeath 2008). Previous studies identified approximately 50 individual star-forming sites. Detected sources are deeply embedded within the molecular filaments and most of these detected sources are identified as prestellar cores or protostellar cores (e.g., Chini et al. 1997; Lis et al. 1998; Johnstone \& Bally 1999; Nielbock et al. 2003; Takahashi et al. 2006; 2008a, b; Peterson \& Megeath 2008).

With interferometric continuum observations, Matthews et al. (2005) and Takahashi et al. (2009) found a bright continuum peak (hereafter \hbox{MMS 6-main}), which  has at least a factor of five larger flux density at (sub)millimeter wavelengths as compared to all the other \hbox{OMC-2/3} sources (Takahashi et al. 2009). However, no star formation signatures such as a clear CO outflow, a radio jet, or an infrared source \hbox{(${\leq}$8 $\mu$m)}, have been detected at the source center in the previous studies (Matthews et al. 2005; Takahashi et al. 2009). The core density profile suggests a power-law structure with an index of -2.0.  This power-law index suggests that the core still retains the density structure of the prestellar phase (Takahashi et al. 2009).  The bolometric luminosity and core mass of MMS 6, as derived from the previous single-dish millimeter to far-infrared data, are \hbox{$<$60 L$_{\odot}$} and \hbox{36 M$_{\odot}$}, respectively (Chini et al. 1997).  These parameters are approximately one order of magnitude higher than the values derived in typical low-mass star-forming regions such as the Taurus molecular cloud (c.f., Myers \& Benson 1983). Therefore, MMS 6 would be one of the first cases for the detailed study of the intermediate-mass star forming core at the earliest evolutionary stage (c.f., Beltran et al. 2008; Fuente et al. 2007; Takahashi et al. 2009; Enoch et al. 2009).

In order to study the physical properties and the evolutionary stage of \hbox{MMS 6-main} more accurately, we have obtained the highest angular resolution image possible with the SMA, utilizing all the configurations at  \hbox{850 $\mu$m}. Our observations have achieved approximately 10 times better angular resolution in terms of beam area, as compared to that of the previous SMA observations by Takahashi et al. (2009).  We compare our results with the theoretically predicted star forming core models such as the prestellar core
model, the protostellar core plus central disk model, and the first adiabatic core model. We make our comparisons in the visibility-amplitude domain and also in terms of the spectral energy distributions (SEDs).

We describe the observations in \hbox{Section 2}. Observational results including sub-arcsecond images, physical properties, and core internal structures, will be presented in \hbox{Section 3}.  Comparisons between the observed core density structures and the theoretical models, as well as the evolutionary stage of MMS 6-main, will be discussed in Section 4. Section 5 summarizes the paper.

\section{OBSERVATIONS AND DATA REDUCTION}
The observations have been performed with the SMA (Ho et al. 2004), utilizing the sub-compact, compact, extended, and very extended configurations.  The \hbox{850 $\mu$m} continuum data were taken with seven antennas in the extended configuration (EX; projected baseline ranging between 40 and \hbox{240 k$\lambda$}), and the very extended configuration (VEX; projected baseline ranging between 90 and \hbox{580 k$\lambda$}) on February 9 and September 2, 2010, respectively. The typical system noise temperatures in DSB mode were between \hbox{200--350 K} for both the EX and VEX configurations at the observed elevations.  Both the LSB and USB data were obtained simultaneously with the \hbox{90$^{\circ}$} phase switching technique by the digital spectral correlator, which has a bandwidth of \hbox{4 GHz} in each sideband. After subtracting the channels where molecular lines might be expected within the band, the LSB and USB continuum data separated by \hbox{10 GHz} RF were combined to improve the sensitivity.  The effective bandwidth for the continuum emission is approximately \hbox{7.5 GHz}.  The phase and amplitude calibrator, 0423-013 (\hbox{1.5 Jy} for EX and \hbox{2.9 Jy} for VEX), was observed every \hbox{18 minutes}.  Observations of Callisto (for EX) and Vesta (for VEX) provided the absolute scale for the flux density calibrations.      The overall flux uncertainty was estimated to be \hbox{$\sim$20\%}.  The passband across the bandwidth was determined from observations of \hbox{3C 454.3} with a \hbox{$\sim$30 minute} integration. The EX and VEX configuration data were combined with the similar wavelength continuum data taken with the SMA sub-compact and compact configurations. Detailed observing settings of the sub-compact and compact configurations are described in Takahashi et al. (2009; 2012).

The raw data were calibrated using MIR, originally developed for the Owens Valley Radio Observatory (Scoville et al. 1993) and adopted for the SMA. After the calibration, the data from all the configurations were combined using the AIPS task \hbox{``DBCON''}, and final CLEANed images were made using the DIFMAP task. In order to reduce residual phase errors and improve the dynamic range of
the images, self-calibrations have been applied. This improved the maximum dynamic range by a factor of 4.  The resulting synthesized beam sizes of the
combined data set were \hbox{0$''$.35$\times$0$''$.30} (or \hbox{145 AU $\times$ 124 AU} at the adopted distance of \hbox{414 pc}) with a position angle of \hbox{32$^{\circ}$} for the uniform weighting image, and \hbox{1$''$.9$\times$1$''$.7} (or  \hbox{787 AU $\times$ 704 AU}) with a position angle of \hbox{76$^{\circ}$} for the Gaussian tapered image \hbox{(FWHM=60 k$\lambda$)}. The achieved rms noise levels were \hbox{2.7 mJy beam$^{-1}$} for the uniform weighting image and \hbox{12 mJy beam$^{-1}$} for the Gaussian tapered image.  Note that the self-calibrated images lose their absolute positions. The positional offsets between the non-self-calibrated image and the self-calibrated images (i.e., \hbox{Figure 1$a$, $b$}) are ${\sigma_{\rm{SL}}}{\sim}$0$''$.048, and 0$''$.12 for the uniformly weighted and the tapered images.  

In addition, the positional accuracy is mostly determined by the phase drift due to the baseline error, (${\sigma_{\rm{BS}}}{{\approx}0.1{\lambda}}{\approx}$0$''$.1) of the non-self-calibrated image \footnote{The baseline error is also measured using the primary calibrator (0423-013) against the secondary calibrator (0530+135).  Both calibrators are located ${\sim}18^{\circ}$ from MMS 6-main, and the separation between two calibrators is $\sim$23$^{\circ}$. The positional shift from the phase center of the secondary calibrator was measured to be ${\sim}0''.1$,
which is consistent with the expected baseline error of the SMA, 0.1$\lambda$.}.
Finally the positional accuracies of the final CLEANed images are estimated to be ${\sigma}_{\rm{total}}{\approx}{\sqrt{{\sigma_{\rm{BS}}}^2+{\sigma_{\rm{SL}}}^2}}{\approx}0''.11$ for the uniform weighted image and ${\approx}0''.16$ with the Gaussian tapered image.

\section{RESULTS}
\subsection{Subarc-second 850 $\mu$m Continuum Image}
Figure 1 $a$ and $b$ show the \hbox{850 $\mu$m} continuum images, which are predominantly thermal dust emission (Takahashi et al. 2009), obtained with the SMA.  As denoted by crosses in \hbox{Figure 1$a$}, five infrared sources
(i.e., IRS1--5; protostellar and young stellar objects) are located around \hbox{MMS 6-main}, but none of them are associated with the peak position of the \hbox{850 $\mu$m} continuum emission of \hbox{MMS 6-main}.  Note that the positional shift between MMS 6-main and the closest infrared source, IRS 3, is 2$''$. 
As discussed in Section 3.1.2 in Takahashi et al. (2009), the accuracy of the positional alignment between the infrared image and 850 $\mu$m image is on the order of 0$''$.3, which was estimated from a common compact continuum source obtained in the same region and during the similar observational period. In addition, the positional accuracy of MMS 6-main obtained with the SMA 850 $\mu$m (uniformly weighted image) is 0$''$.11. These comparison show that the positional shift between MMS 6-main and IRS 3 is most likely real.

In order to derive the peak flux, flux density, and source size, a 2D-Gaussian fit with two components, was applied to the MMS 6-main map. The derived properties for a compact and an extended component, are reported in Table 1. The residuals after the two-component fit are the 20\% level at most as compared to the peak flux. The non-uniform gas distribution, with substructures as described in the later part of this Section as well as non-Gaussian shaped envelope structures, are the causes of the residuals.  The total flux and the peak intensity of the compact component are measured as 1.1 Jy and 0.56 Jy beam$^{-1}$, respectively.  The brightness temperature is estimated to be 52 K from the peak intensity within the beam size. The gas temperature is equal to the brightness temperature when the gas is optically thick \hbox{($T_b=T_{\rm{gas}}$ for ${\tau}{\geq}1$)}, while the gas temperature is larger than \hbox{52 K} when the gas is optically thin \hbox{($T_b={\tau}T_{\rm{gas}}$ for ${\tau}<<1$)}.  Assuming the gas temperature is \hbox{52 K}, \hbox{$\tau$=1}, the gas-to-dust ratio of 100, the column density is estimated to be \hbox{$N_{H_2}$=(2.0${\pm}$0.4)${\times}$10$^{25}$ cm$^{-2}$}. This gives us the source mass of \hbox{M=0.28$\pm$0.06 M$_{\odot}$}, and the mean number density of \hbox{$n_{H_2}$=1.4${\pm}$0.3${\times}$10$^{10}$ cm$^{-3}$}, with the assumption of a spherically symmetric structure \footnote{Fitting errors of the deconvolved size and flux as well as the 20\% absolute flux uncertainty described in Section 2 were considered in order to estimate the error of physical parameters.}.
For these estimates, the dust emissivity index of \hbox{$\beta$=0.93} (Takahashi et al. 2009) and the dust absorption coefficient of \hbox{$\kappa_{\lambda}$=0.027 cm$^{2}$ g$^{-1}({\lambda}$/400 $\mu$m)$^{-{\beta}}$} are  assumed (Keene et al. 1982).

The total SMA flux density of the compact continuum component associated with MMS 6-main corresponds to 17 \% of the flux density measured within the JCMT/SCUBA 13$''$ beam at \hbox{850 $\mu$m} (6.6 Jy; Johnstone \& Bally 1999).  Note that the ratio in beam surface area between the JCMT and the SMA is \hbox{$\sim$1900}.  A large amount of flux at the submillimeter wavelength, is concentrated within the central \hbox{100 AU} size scale.

In addition to the \hbox{MMS 6-main} component, a number of substructures extending from the \hbox{MMS 6-main} peak, have been spatially resolved in our continuum image (denoted as dashed blue arrows in Figure 1c and Figure 2).
Here, we identify ``substructures'' as those distinct features whose peak fluxes have a signal-to-noise ratio of more than \hbox{25$\sigma$}, and whose contrasts with respect to their fainter surroundings are more than \hbox{$10{\sigma}$}.
We argue that these substructures, at the level of 10 times fainter than the MMS 6-main peak, are not artifacts due to sampling or image reconstruction.  This is verified by calibrating and imaging our secondary calibrator, 0530+135, in exactly the same manner as for MMS 6. 
In this image, a dynamic range of at least 50 is achieved, with no hint of any substructure for the secondary calibrator, down to the theoretical noise level.  We have also compared the maps made with uniform or natural weighting,
and with or without self-calibration. The same structures are seen with good correspondence between the different maps.   Hence we conclude that the detected substructures are robust.

Among the detected substructures, four are particularly well resolved. We mark them with orange circles in Figure 1c and Figure 2.  Their contrast with the surrounding structures at the same radii is much greater than 10 sigma.  We call these four, the ``sub-clumps''. All these sub-clumps are located at 0$''.$9--1$''$.7 (370-700 AU) from the MMS 6-main peak position. The sizes of the sub-clumps are comparable to the SMA synthesized beam.

The peak fluxes of the sub-clumps are \hbox{70--80 mJy beam$^{-1}$}. Assuming that the dust emission is optically thin within these sub-structures, and that the temperature distribution of the dust continuum emission is uniform with \hbox{$T_d$=20 K}, the masses of the sub-clumps are estimated to be \hbox{M=0.066 -- 0.073 M$_{\odot}$}.  Such masses are on the order of the values for brown dwarfs,  (0.011${\leq}{M_{\ast}}{\leq}$0.084 M$_{\odot}$).
Assuming that the sub-clumps have a similar size as the synthesized beam and a spherical geometry, the number densities are estimated to be \hbox{$n_{H_2}{\approx}3{\times}$10$^{9}$ cm$^{-3}$}.
\hbox{Figure 2} shows the flux distribution along the azimuthal direction as a function of radius from the \hbox{MMS 6-main} peak position. The flux fluctuations along the azimuthal directions, are exactly what we had noted in \hbox{Figure 1c}.  Separations between these sub-clumps/spikes range between
\hbox{0.6-1.4$''$} \hbox{(250--580 AU}; as denoted by arrows in \hbox{Figure 2}). The origin of these sub-structures is discussed in \hbox{Section 4.3}.

In addition, extended emission at the \hbox{5$\sigma$} level, is detected to the southeast of \hbox{MMS 6-main}. This elongation is consistent with the large scale filamentary structure (e.g., Chini et al. 1997), and may reflect the fragmentation processes within the filaments.

\subsection{Visibility-Amplitude Plot}
The high flux density concentration is attributed to not only the density distribution, but also to the temperature distribution. In order to define these distributions closer to the center, we further resolve the core internal structure as first elucidated by Takahashi et al. (2009). We use the newly obtained SMA EX and VEX configuration data to study approximately one order
of magnitude smaller size scales \hbox{(${R}\sim$60 AU)} as compared to the previous analysis.
Core internal structures are discussed in the Fourier domain analysis by utilizing the visibility-amplitude plot as presented in \hbox{Figure 3}. We directly analyze the interferometric data, bypassing the non-linear image deconvolution process and the problem of the missing flux.

Note that the binned amplitude-visibility plot is processed with the vector amplitude average. Substructures shifted from the phase center of the visibility plot will be attenuated due to the coherence loss. This effect suppresses the flux contamination from the substructures. Asymmetric structures within the core are presented as a variation of the visibility-amplitudes as a function of the projected $uv$-distance.

We assume that the density and temperature follow simple radial power laws as
\hbox{${\rho}(r){\propto}r^{-p}$} and \hbox{$T(r){\propto}r^{-q}$}, respectively.  Then the power law of the visibility amplitude plot, $\gamma$, have a relation of \hbox{${\gamma}={p}+{q}-3$} (e.g., Harvey et al. 2003a; Takahashi et al. 2009).  Figure 3 shows the visibility amplitude plot as a function of the $uv$ distance. Our observational results suggest two different power-law indices, with the break at \hbox{60 k$\lambda$} \hbox{($D=$1300 AU)}.
The outer part and inner part of core are fitted by $\gamma$=-0.64 and -1.4, respectively.
In the isothermal case ($q$=0), the density power-law index of the outer and inner part of core are derived to be 2.4 and 1.6, respectively. Assuming the central heating source ($q$=0.41 derived by Takahashi et al. 2009; optically thin case), the density power-law index of the outer and inner part of core are derived to be 2.0 and 1.2, respectively.  These results suggest that the density profile is shallow at the inner part of core \hbox{($D{\leq}$1300 AU)}.

\section{DISCUSSION}
\subsection{Model Comparisons}
In order to further discuss the internal structure, the observed results are compared with core models in the different evolutionary stages based on the visibility-amplitude(\hbox{Figure 3}) and the spectral energy distribution (Figure 4): (i) Prestellar core, (ii) First adiabatic core, and (iii) Protostellar core. For all models, the gas temperature of 20 K, which is the typical gas temperature derived from the northern part of OMC clumps, is adopted (Cesaroni \& Wilson 1994). Moreover, an effective sound speed of \hbox{0.42 km s$^{-1}$}, which is derived from the \hbox{H$^{13}$CO$^{+}$} observations (Takahashi et al. 2009), was included in order to take account of the observed non-thermal motions.  The radiative transfer calculations are performed to produce the visibility-amplitude plot and SED for model cores. For the calculations, composite aggregates dust model (Semenov et al. 2003) was adopted.

{\bf (i) Prestellar core:} Larson-Penston model ($c_{\rm{eff}}$=0.42 km s$^{-1}$ and $t=0$; the end of the prestellar phase) is adopted (denoted by a blue curve in \hbox{Figure 3}). Note that the Larson-Penston model is dependent only on the sound speed.
The measured sound speed of 0.4 km s$^{-1}$ by Takahashi et al. (2009) was adopted. 
The simple prestellar core model shows approximately one order of magnitude smaller visibility amplitude distribution as compared with the observational result. 
The results unlikely show that MMS 6-main is the prestellar core. Recent discovery of an extremely compact molecular outflow associated with MMS 6-main by Takahashi \& Ho (2012) 
supports to discard the prestellar scenario. 

{\bf (ii) The first adiabatic core:}  This model, denoted by a green curve in Figure 3, is calculated using the three dimensional radiative hydrodynamic numerical simulations  (e.g., Tomida et al. 2010b; Saigo et al. 2012).
In order to investigate the formation of the intermediate mass star, we adopt the initial condition of the rotating critical Bonnor-Ebert sphere with the effective sound speed of \hbox{$c_{\rm{eff}}=$0.42 km s$^{-1}$}, which was measured from the H$^{13}$CO$^{+}$ (1--0) observations (Takahashi et al. 2009).
The central density of \hbox{$n_{H_2}$=2.6$\times$10$^{4}$ cm$^{-3}$}, which is a typical density for nearby molecular clouds as well as the value usually used for the first adiabatic core simulations (e.g., Saigo et al.2008; 2011), was adopted as an initial density for the parental molecular cloud. In this calculation, we assume the gas temperature of 20 K.  
The age of the plotted first adiabatic core is $\approx$2000 yr. The plotted model is the best model to reproduce observed flux density among their evolution. 
Resulting visibility amplitude distribution are denoted by the green curve in \hbox{Figure 3}. The shape of the visibility distribution is not consistent with the observed data. 
Note that younger phase of the simulated first adiabatic core models show lower flux values than presented model due to less significant gravitational potential of the first core, 
while later phase of the first adiabatic core models does not show much differences in terms of the total flux, but show even flatter visibility amplitude distributions for much of the baseline range compared to the presented model.

{\bf (iii) Protostellar core:} This model assumes the Larson-Penston core + central disk structure (denoted by a red curve in \hbox{Figure 3}).
Parameters of the protostellar model are selected for reproducing the observed amplitude.  For the disk-like structure, \hbox{$R=300$ AU}, \hbox{$M=1.5$ M$_{\odot}$; ${\Sigma}(r){\propto}r^{-1}$}, and \hbox{$T$=30 K} are assumed.
Moreover, the temperature distribution in the envelope of the central protostar was assumed as \hbox{$T(r)=500(r/10$ AU)$^{-1}$ K}. The given temperature gradient, \hbox{$T(r){\propto}r^{-1}$}, is considering a central heating source and an optically thick envelope (Hartmann 1998).

The prestellar core model has a lower flux as compared with the other two models. This is due to the cool gas temperature \hbox{(20 K)} as expected in a prestellar core. The observed large flux at \hbox{850 $\mu$m} is not consistent with the prestellar core model, but clearly suggests that a self-luminous source (i.e. higher temperature gas than \hbox{20 K}) is required to explain the observed large flux at \hbox{850 $\mu$m}.

Note that the visibility amplitudes present a bump or excess between \hbox{200 k$\lambda$} and \hbox{400 k$\lambda$} as denoted by the blue arrows in \hbox{Figure 3} as denoted by a blue arrow. 
However, the bump observed at $\approx$200 k$\lambda$ corresponds to the linear size of \hbox{$R{\approx}$300 AU}.
We can produce such a bump in the protostellar core by adding a \hbox{30 K}, 300 AU disk-like structure as we shown in protostellar model in Figure 3. The addition of such a disk is too simplistic, in view of the observed structures.  The observed bump in the visibility amplitudes above a smooth extended core, is due to the clumpy structures including the MMS 6-main peak which we have detected in Figure 1c, and which all have a characteristic size scale of around 300 AU. Furthermore, the visibility amplitude variations are a function of the azimuth angle (\hbox{Figure 2}).

Figure 4 presents the comparisons between the observed \hbox{MMS 6-main} spectral energy distribution and their core models. Black open squares in \hbox{Figure 4} show the observed SED of \hbox{MMS 6-main} measured with an angular resolution of \hbox{$0''.6$--$6.''0$}. These values were retrieved from \hbox{Table 5} in Takahashi et al. (2009). Spitzer/MIPS \hbox{24 $\mu$m} and \hbox{70 $\mu$m} upper limits are also shown in the SED. These upper limits are due to the contamination from another \hbox{Class I} source, \hbox{IRS 3}, located at \hbox{2$''$} north of \hbox{MMS 6-main}. On the other hand, \hbox{850 $\mu$m} continuum emission obtained with the SMA, \hbox{2.3 mm} and \hbox{3.3 mm} continuum data obtained with the Nobeyama Millimeter Array, and \hbox{7.3 mm} continuum emission obtained with the Very Large Array, are from lower limits due to the missing flux. However, there are no contamination from nearby sources at these wavelengths because of the high-angular resolution observations.

Observed millimeter/submillimeter continuum emission have higher flux densities as compared with the SED expected in the prestellar and first adiabatic core models denoted by the blue line and the green line in \hbox{Figure 4}. For example, approximately a factor of six higher flux density was observed at \hbox{850 $\mu$m} as compared to these two models. The protostellar model as denoted by the red line agrees well with the observed flux densities.  SED comparisons clearly imply that the observed high flux densities are caused by a self-luminous source.

\subsection{Evolutionary Stage of MMS 6-main}

In terms of the core, the sizes of \hbox{MMS 6-main} and the other protostellar sources in the \hbox{OMC-3} region do not vary significantly (i.e., core sizes measured at more than \hbox{5$\sigma$} signal level at \hbox{850 $\mu$m} range between \hbox{900 AU} and \hbox{2100 AU}; Takahashi et al. 2012). This suggests that \hbox{MMS 6-main} may have a similar density structure in the outer most region. However, \hbox{MMS 6-main} has more than a factor of five larger millimeter/submillimeter flux densities as compared with other sources detected in \hbox{OMC-2/3} with any aperture size between \hbox{10000 AU} and \hbox{500 AU} (e.g., Chini et al. 1997; Takahashi et al. 2009; 2012).

The visibility amplitude plot and SED clearly imply the presence of a central heating source embedded within \hbox{MMS 6-main}. Hence, \hbox{MMS 6-main} is most likely in the protostellar core phase. The absence of a mid-infrared source, a large-scale molecular outflow, or an ionized jet, with current sensitivity and angular resolution, can be explained if the central driving
source is extremely young (less luminous with high-extinction from surrounding circumstellar material), so that the outflow and jet are not yet substantial.
Due to the high-extinction, the infrared emission is absorbed by the massive surrounding envelope and then reradiated in the submillimeter wavelengths.
Takahashi \& Ho (2012) have recently discovered an extremely compact molecular outflow (1000 AU) associated with \hbox{MMS 6-main} in \hbox{CO (3--2)} and \hbox{HCN (4--3)}. This is the smallest molecular outflow associated with the intermediate-mass protostars and also even smaller than the outflows found
in the protostellar cores in the Taurus star forming region (e.g., Hogerheijde et al. 1998). Nevertheless, the estimated outflow force (${\sim}10^{-4}$ M$_{\odot}$ km s$^{-1}$ yr$^{-1}$) is similar as compared to the other large-scale molecular outflows associated with the protostellar cores in the OMC-2/3 region. Furthermore, the observed bow-shock type of outflow velocity structures are similar to those observed in the Class 0 type protostars. These results support our suggestion that \hbox{MMS 6-main} is in the protostellar phase, and in particular at a very early evolutionary phase.

\subsection{Substructure within the Massive Core}
Substructures were detected in our sub arcsecond resolution of 850 $\mu$m continuum image for the first time. 
There are two possible formation mechanisms to produce the substructures and sub-clumps surrounding \hbox{MMS 6-main}:
{\bf (i) gravitational instability within dynamically collapsing cores, or (ii) gravitational instability within rotationally supported disks.}

{\bf (i) dynamically collapsing core case:}
Hanawa \& Matsumoto (2000) suggest that collapsing cloud cores become more unstable if the effective sound speed (e.g., turbulent motions) decreases during the collapse (see also Saigo et al. 2000). Non-thermal velocity width of \hbox{0.42 km s$^{-1}$} \hbox{(mach=2.5)} are observed in the outer part of \hbox{MMS 6} core \hbox{($\sim$1500 AU)}.  The Jeans length is described as \hbox{${\lambda}_{\rm{frag}}={\sqrt{{\pi}{c_s^2}/G{\rho}_0}}$}. Here, $c_s$ is the sound speed, and the relation between the sound speed and the temperature of
gas ($T_d$) is described as $c_s=\sqrt{kT_d/{\mu}m_H}$. G, ${{\rho}_0}(={\mu}m_Hn_{H_2})$, $\mu$, and $m_H$ are the gravitational constant, the mean density of the gas, and the hydrogen mean molecular weight of 2.33, and the hydrogen mass, respectively.  Note that both the temperature and density vary slowly and in the same sense, so that the Jeans analysis is approximately correct as an estimate. Adopting the gas temperature of \hbox{20 K}, and the typical number density of \hbox{${n_{H_2}}\simeq$3.0$\times$10$^{9}$ cm$^{-3}$} (derived in Section 3.1), the Jeans length is estimated to be \hbox{360 AU (0.$''$87)}. Note that the number density derived from the sub-clumps is used to represent the number density of the parental gas (i.e., the number density before the fragmentation).  Hence, the estimated Jeans length is the lower-limit of the fragmentation scale, if fragmentation occurred much earlier. The derived Jeans length is similar to the separation between the sub-clumps (i.e., the  separations of local peaks as measured in \hbox{Figure 2}),  which is \hbox{0.6$''$--1.4$''$} \hbox{(250--580 AU)}.

{\bf (ii) rotationally supported disk:} Shu et al. (1990) show that if the rotationally supported disk becomes sufficiently massive,
\hbox{$M_{d}/(M_{\ast}+M_{\rm{dis}}){\sim}$24\%}, the rotationally-supported disk becomes gravitationally unstable and spiral arms are formed. In the MMS 6-main case, a 2.5 M$_{\odot}$ disk, which corresponds to  \hbox{$M_{d}/(M_{\ast}+M_{\rm{disk}}){\sim}$45\%}
\footnote{Here, stellar mass of 3 M$_{\odot}$ was assumed as a maximum stellar mass at this stage. Mass of the disk-like structure was estimated from the total flux density detected at more than 5$\sigma$ level (4.7 Jy), after subtracting the flux contribution from MMS 6-main (2.0 Jy) with the assumption of \hbox{$T_d$=20 K}}, might be suggested from the observations.
In such a case, it is possible that the observed sub-structures could be due to instability within the purported disk. Fragmentation of self-gravitating massive disk-like structure has also been suggested by other studies (e.g., Gammie 2001; Kratter \& Matzner 2006; Kratter et al. 2008). We emphasize that currently, our observations only indicate substructures, without any definitive evidences of a disk-like structure.  With current sensitivity, we also do not have any kinematic information from the envelope/disk-like structures.  Sensitive molecular line observations are required to investigate the kinematics and to test the proposed scenario.

In both cases, the physical processes follow the self-gravitational forces.
Moreover, it is important to note that fragmentation within the disk will disturb the mass accretion processes onto the central star (e.g., Kratter et al. 2006; 2008). This process plays an important role in determining the final mass of the primary star.  Our observations are the first case to present fragmented sub-structures within core/disk-like structure around an embedded protostar with $R{\approx}$500 AU.  As we mentioned in Section 3.1, the mass of sub-clumps (0.066-0.073 M$_{\odot}$) have brown dwarf mass.  Origin of the detected sub-clumps could be the eventual cluster members around the primary star or related to brown dwarf/planet formation.

\subsection{Future Prospects}
Limited data quality at the projected $uv$ distances of longer than \hbox{400 k$\lambda$}, as well as the presumably optically thick continuum emission at \hbox{850 $\mu$m}, make it difficult to constrain the detailed  spatial structures within \hbox{$D{\leq}$100 AU} scale. Even higher angular resolution observations with the relatively optically thin wavelengths
\hbox{(${\lambda}{\geq}850~{\mu}$m )} will be necessary for defining the detailed internal density and temperature structures. Observations with the EVLA (e.g, K, Ka, and Q band) and ALMA \hbox{(e.g., Band3--6)} will be a key for these studies. These will also reveal detailed structures within the spikes and sub-clumps and observationally constrain the origin of these structures.
Observations using the high density tracer or hot core tracer with the SMA, ALMA and EVLA are important for revealing gas physical conditions and dynamics of the inner envelope, as well as the circumstellar disk (i.e., rotationally supported disk), if present, at the earliest evolutionary phase (e.g., Zapata et al. 2010). Moreover, high angular resolution \hbox{($<<1''$)} and high sensitivity EVLA observations are also crucial in order to search for the mass ejection phenomena associated with the central star such as faint ionized jet and/or \hbox{H$_2$O} maser.

\section{Conclusion}

With the highest angular resolution possible with the SMA, an intermediate-mass protosatar, MMS 6-main, has been spatially resolved. The observed flux density at the MMS 6-main ($D{\approx}$120 AU) corresponds to the brightness temperature of \hbox{52 K}. The column density and the mass are estimated to be \hbox{$N_{H_2}=$2.1$\times$10$^{25}$ cm$^{-2}$} and \hbox{$M$=0.29 M$_{\odot}$}, respectively. Detailed model comparisons clearly show that the observed visibility amplitudes cannot be explained by the simple prestellar core model with a gas temperature of \hbox{20 K}. A self-luminous source is necessary to explain the observed high-visibility amplitudes. Multi-wavelength comparisons using the SED as well as recent extremely compact molecular outflow detection by Takahashi \& Ho (2012) also support a protostellar type source. Several sub-clumps and spikes have been spatially resolved. These structures are well centered at \hbox{MMS 6-main}. Masses of the detected sub-clumps are estimated to be on the order of brown dwarf mass \hbox{(0.066--0.073 M$_{\odot}$)}.
The origin of the density fluctuations can be explained by the gravitational instability within the massive core or a massive disk-like structure.
These structures could form brown dwarfs or planets around the primary star.

\acknowledgments
We thank our two anonymous referees for very helpful suggestions which improved the clarity and logic of the paper.  We acknowledge the staff at the Submillimeter Array for their assistance with the observations.
We thank Keiichi Asada for fruitful discussions.
S.T. is financially supported by a postdoctoral fellowship at the Institute of Astronomy and Astrophysics, Academia Sinica, Taiwan.
K.T. is supported by the Research Fellowship from the Japan Society for the Promotion of Science (JSPS) for Young Scientists.

\begin{deluxetable}{llllll}
\tablecaption{Parameteres of MMS 6-main\label{}}
\tablewidth{0pt}
\tablehead{
\colhead{} & \colhead{RA}  & \colhead{Dec} & \colhead{Deconvolved Size, P.A.} & \colhead{Peak Intensity} &\colhead{Flux Density} \\
\colhead{} & \colhead{[J2000]} & \colhead{[J2000]} & \colhead{[arcsec, degree]} & \colhead{[Jy beam$^{-1}$]} & \colhead{[Jy]} \\
}
\startdata
Compact    & 05 35 23.42 & -05 01 30.57  & 0.318$\pm$0.004${\times}$0.295$\pm$0.004 (174$\pm$7$^{\circ}$)   & 0.556${\pm}$0.003 & 1.061${\pm}$0.008 \\
Extended   & 05 35 23.41 & -05 01 30.57  & 1.788$\pm$0.001${\times}$1.448$\pm$0.001 (156$\pm$2$^{\circ}$)   & 0.152${\pm}$0.002 & 3.931${\pm}$0.045  \\
\enddata
\end{deluxetable}

\begin{figure*}[t]
       \epsscale{1.0}
       \plotone{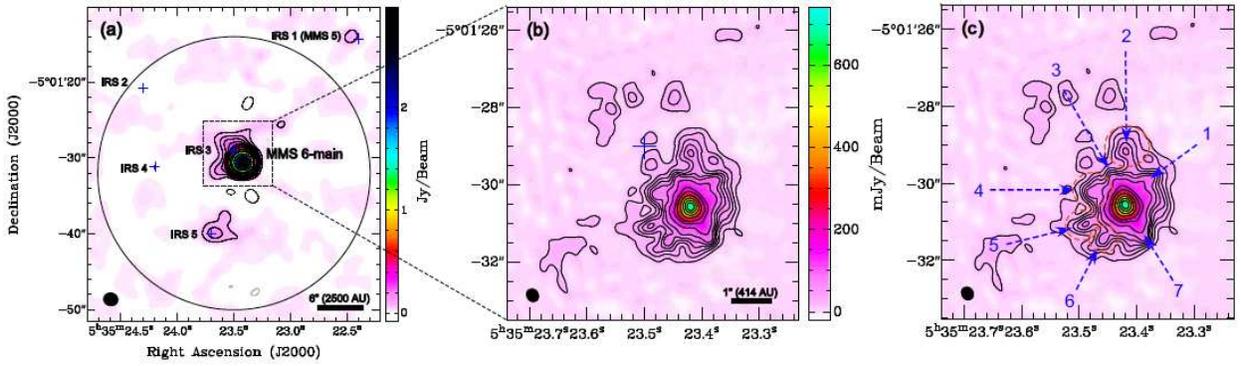}
     \caption{The 850 $\mu$m continuum images taken with the SMA subcompact, compact, extended, and very
     extended configurations. {\it (a)}: Image made with $uv$
     tapering of FWHM=60 k$\lambda$. The contour levels are ${\pm}$5$\sigma$, 10$\sigma$, 15$\sigma$, 20$\sigma$, 25$\sigma$, 30$\sigma$,
     40$\sigma$, 60$\sigma$, 80$\sigma$, 120$\sigma$, 160$\sigma$, 200$\sigma$, and 240$\sigma$ (1$\sigma$=12 mJy beam$^{-1}$).
     {\it (b)}: Image made with uniform weighting. The contour levels are ${\pm}$5$\sigma$, 10$\sigma$, 15$\sigma$, 20$\sigma$,
     25$\sigma$, 30$\sigma$,40$\sigma$, 60$\sigma$, 80$\sigma$, 120$\sigma$, 160$\sigma$, 200$\sigma$, and 240$\sigma$ (1$\sigma$=2.7 mJy beam$^{-1}$).
     {\it (c)}: Same as figure (b), overlaid with notes for detected substructures as denoted by blue dashed arrows
     and the sub-clumps as denoted by orange dashed circles.
     Denoted numbers for each substructure are corresponding to same numbers as in Figure 2.
     Negative contours are presented by the dashed lines. The crosses show the positions of the infrared sources identified in Takahashi et al. (2009).
     The size of the crosses corresponds to the positional accuracies of the Spitzer images in Takahashi et al. (2009).
     Filled ellipses at the bottom-left corners show the synthesized beams of each image.
     These figures were made using the KVIS in KARMA (Gooch 1995). \label{}}
\end{figure*}

\begin{figure*}[t]
       \epsscale{0.8}
       \plotone{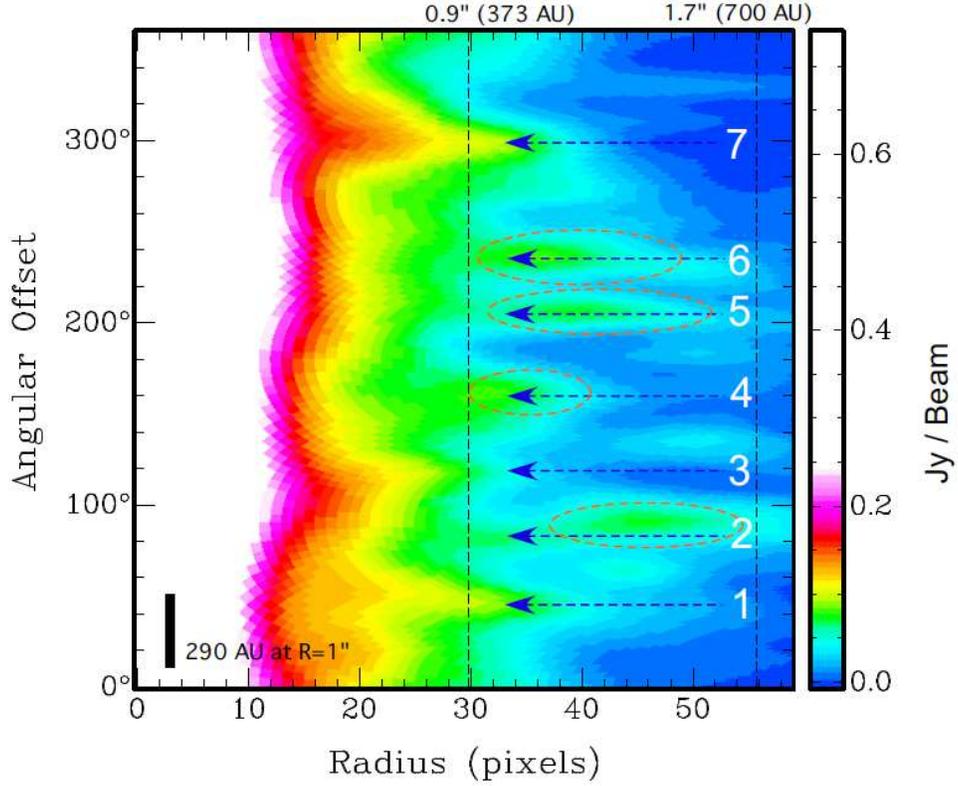}
     \caption{\emph{The flux distribution as a function of azimuthal direction and radius centered at the peak position of MMS 6-main.
     1 pixel corresponds to 0$''$.03. The black lines in the vertical axis show 1$''$ and 1.$''$5 from the center.
     Positions of spiky substructures are denoted by blue dashed allows. Sub-clumps denoted by dashed orange circles.
     Denoted numbers of each substructure are corresponding to same numbers denoted in Figure 1c.
     East direction is defined as the angular offset as 0$^{\circ}$ and is counted toward counter clock direction as a positive offset.
     Figure was made using the KPOLAR in KARMA (Gooch 1995).
     \label{}}}
\end{figure*}

\begin{figure*}[t]
       \epsscale{0.70}
       \plotone{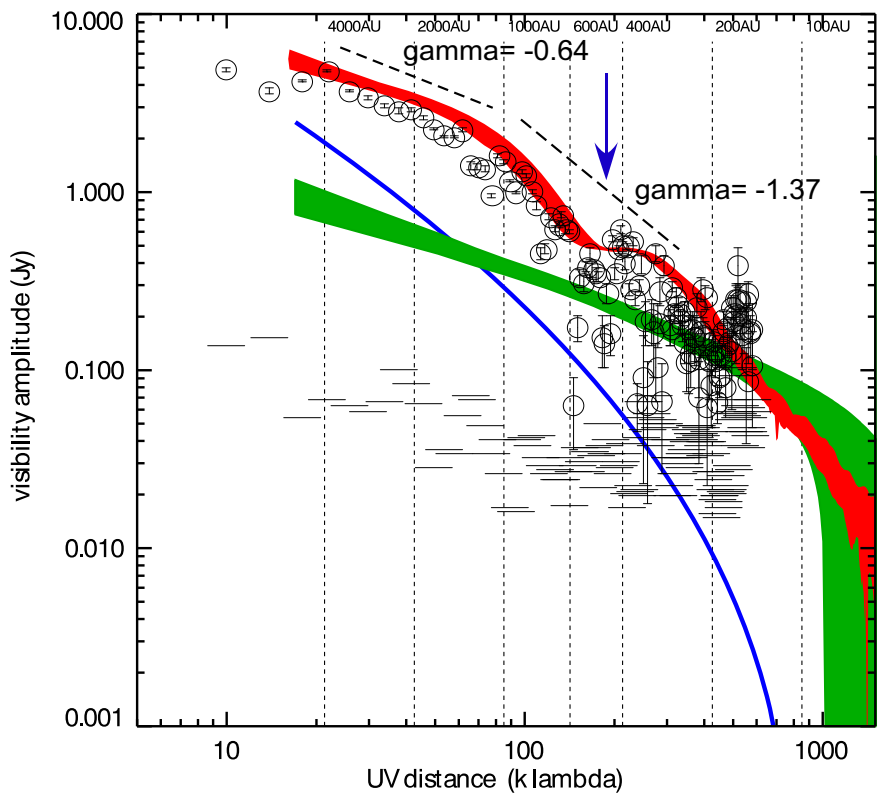}
      \caption{\emph{Visibility amplitude plot binned by 4 k$\lambda$ as a function of the projected $uv$ distance
     produced from the 850 $\mu$m continuum data.
     Black open circles show the observational results obtained from the SMA 850 $\mu$m continuum image.
     The statistical errors (1$\sigma$) of the 850 $\mu$m data are denoted by vertical error bars.
     For the comparisons, prestellar core model (Larson-Penston core) with 20 K gas is denoted by blue curve.
     First adiabatic core model is denoted by green curve.
     Protostellar core model (Larson-Penston core with disk-like structure) is denoted by red curve.
     The core models are calculated with the inclination angle of 30$^{\circ}$.
     \label{}}}
\end{figure*}

\begin{figure*}[t]
       \epsscale{0.70}
       \plotone{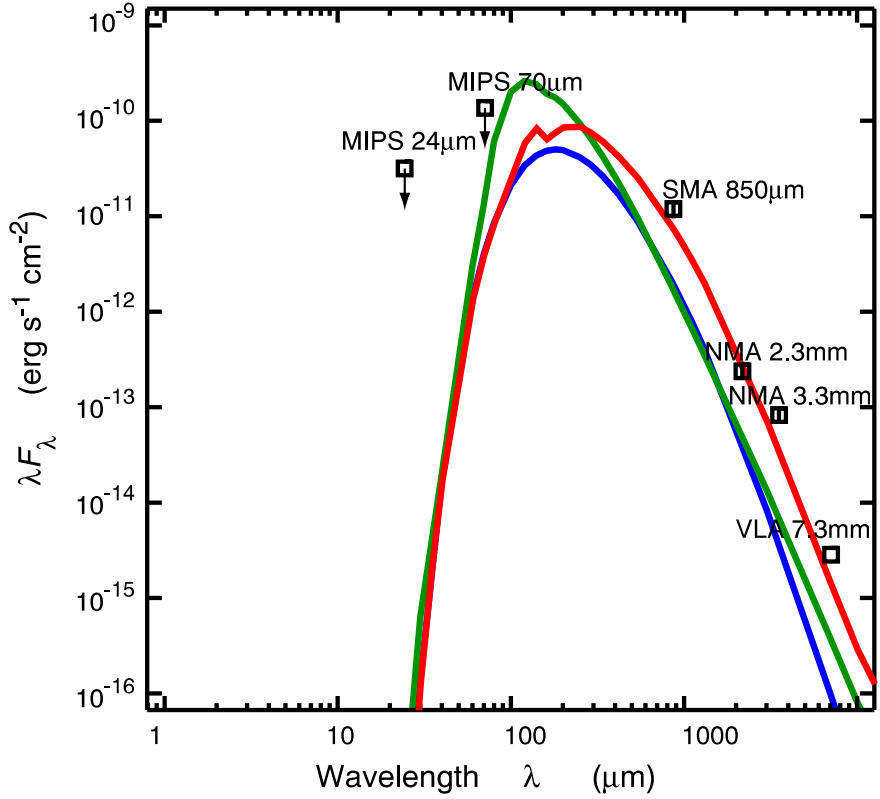}
      \caption{\emph{Spectral energy distributions measured in the millimeter, submillimeter, and mid-infrared wavelengths
     are denoted by open squares. These data were retrieved from Takahashi et al. (2009).
     The bars within the open squares show the error bar of the absolute flux measurements (i.e., ${\pm}20\%$ for 850 $\mu$m, ${\pm}15\%$ for 2.3 mm and 3.3 mm,
     and a few percent for 7.3 mm). Model SEDs were overlaid with the observed data:
     prestellar core model (Larson-Penston core) with 20 K gas is denoted by blue curve.
     First adiabatic core model is denoted by green curve.
     Protostellar core model (Larson-Penston core with disk-like structure) is denoted by red curve.
     The core models are calculated with the inclination angle of 30$^{\circ}$.
     \label{}}}
\end{figure*}


\begin{thebibliography}{}
\bibitem[Aso et al.(2000)]{2000ApJS..131..465A} Aso, Y., Tatematsu, K., Sekimoto, Y., Nakano, T., Umemoto, T., Koyama, K., \& Yamamoto, S.\ 2000, \apjs, 131, 465
\bibitem[Bate(1998)]{1998ApJ...508L..95B} Bate, M.~R.\ 1998, \apjl, 508, L95
\bibitem[Bate(2010)]{2010MNRAS.404L..79B} Bate, M.~R.\ 2010, \mnras, 404, L79
\bibitem[Beltr{\'a}n et al.(2008)]{2008A&A...481...93B} Beltr{\'a}n, M.~T., Estalella, R., Girart, J.~M., Ho, P.~T.~P., \& Anglada, G.\ 2008, \aap, 481, 93
\bibitem[Bonnell(1994)]{1994MNRAS.269..837B} Bonnell, I.~A.\ 1994, \mnras, 269, 837
\bibitem[Cesaroni \& Wilson(1994)]{1994A&A...281..209C} Cesaroni, R., \& Wilson, T.~L.\ 1994, \aap, 281, 209
\bibitem[Chini et al.(1997)]{1997ApJ...474L.135C} Chini, R., Reipurth, B., Ward-Thompson, D., Bally, J., Nyman, L.-A., Sievers, A., \& Billawala, Y.\ 1997, \apjl, 474, L135
\bibitem[Commer{\c c}on et al.(2010)]{2010A&A...510L...3C} Commer{\c c}on, B., Hennebelle, P., Audit, E., Chabrier, G., \& Teyssier, R.\ 2010, \aap, 510, L3
\bibitem[Enoch et al.(2009)]{2009ApJ...707..103E} Enoch, M.~L., Corder, S., Dunham, M.~M., \& Duch{\^e}ne, G.\ 2009, \apj, 707, 103
\bibitem[Fuente et al.(2007)]{2007A&A...468L..37F} Fuente, A., Ceccarelli, C., Neri, R., et al.\ 2007, \aap, 468, L37
\bibitem[Gammie(2001)]{2001ApJ...553..174G} Gammie, C.~F.\ 2001, \apj, 553, 174
\bibitem[Gooch(1996)]{1996ASPC..101...80G} Gooch, R.\ 1996, Astronomical Data Analysis Software and Systems V, 101, 80
\bibitem[Hanawa \& Matsumoto(2000)]{2000PASJ...52..241H} Hanawa, T., \& Matsumoto, T.\ 2000, \pasj, 52, 241
\bibitem[Hartmann(1998)]{1998apsf.book.....H} Hartmann, L.\ 1998, Accretion processes in star formation / Lee Hartmann.~Cambridge, UK ; New York : Cambridge University Press, 1998.~(Cambridge astrophysics series ; 32)  ISBN 0521435072.,
\bibitem[Harvey et al.(2003)]{2003ApJ...583..809H} Harvey, D.~W.~A., Wilner, D.~J., Myers, P.~C., Tafalla, M., \& Mardones, D.\ 2003, \apj, 583, 809
\bibitem[Harvey et al.(2003)]{2003ApJ...596..383H} Harvey, D.~W.~A., Wilner, D.~J., Myers, P.~C., \& Tafalla, M.\ 2003, \apj, 596, 383
\bibitem[Ho, Moran, \& Lo(2004)]{ho04} Ho, P. T. P., Moran, J., \& Lo, K. Y. 2004, \apjl, 616, L1
\bibitem[Johnstone \& Bally(1999)]{1999ApJ...510L..49J} Johnstone, D., \& Bally, J.\ 1999, \apjl, 510, L49
\bibitem[Keene et al.(1982)]{1982ApJ...252L..11K} Keene, J., Hildebrand, R.~H., \& Whitcomb, S.~E.\ 1982, \apjl, 252, L11
\bibitem[Kratter \& Matzner(2006)]{2006MNRAS.373.1563K} Kratter, K.~M., \& Matzner, C.~D.\ 2006, \mnras, 373, 1563
\bibitem[Kratter et al.(2008)]{2008ApJ...681..375K} Kratter, K.~M., Matzner, C.~D., \& Krumholz, M.~R.\ 2008, \apj, 681, 375
\bibitem[Larson(1969)]{1969MNRAS.145..271L} Larson, R.~B.\ 1969, \mnras, 145, 271
\bibitem[Lis et al.(1998)]{1998ApJ...509..299L} Lis, D.~C., Serabyn, E., Keene, J., Dowell, C.~D., Benford, D.~J., Phillips, T.~G., Hunter, T.~R., \& Wang, N.\ 1998, \apj, 509, 299
\bibitem[Machida et al.(2006)]{2006ApJ...647L.151M} Machida, M.~N., Inutsuka, S.-i., \& Matsumoto, T.\ 2006, \apjl, 647, L151
\bibitem[Masunaga \& Inutsuka(2000)]{2000ApJ...531..350M} Masunaga, H., \& Inutsuka, S.-i.\ 2000, \apj, 531, 350
\bibitem[Matsumoto \& Hanawa(2003)]{2003ApJ...595..913M} Matsumoto, T., \& Hanawa, T.\ 2003, \apj, 595, 913
\bibitem[Menten et al.(2007)]{2007A&A...474..515M} Menten, K.~M., Reid, M.~J., Forbrich, J., \& Brunthaler, A.\ 2007, \aap, 474, 515
\bibitem[Myers \& Benson(1983)]{1983ApJ...266..309M} Myers, P.~C., \& Benson, P.~J.\ 1983, \apj, 266, 309
\bibitem[Nielbock et al.(2003)]{2003A&A...408..245N} Nielbock, M., Chini, R., Muller, S.~A.~H.\ 2003, \aap, 408, 245
\bibitem[Peterson \& Megeath(2008)]{aa} Peterson, D. E.,  \& Megeath, S. T., 2006, Handbook of Star Forming Regions, Volume I, 590
\bibitem[Saigo et al.(2000)]{2000ApJ...531..971S} Saigo, K., Matsumoto, T., \& Hanawa, T.\ 2000, \apj, 531, 971
\bibitem[Saigo et al.(2008)]{2008ApJ...674..997S} Saigo, K., Tomisaka, K., \& Matsumoto, T.\ 2008, \apj, 674, 997
\bibitem[Saigo \& Tomisaka(2011)]{2011ApJ...728...78S} Saigo, K., \& Tomisaka, K.\ 2011, \apj, 728, 78 (Saigo et al. 2011a)
\bibitem[Saigo et. al. (2012)]{2011ApJ...728...78S} Saigo, K. et. al. in preparation
\bibitem[Scoville et al.(1993)]{1993PASP..105.1482S} Scoville, N.~Z., Carlstrom, J.~E., Chandler, C.~J., Phillips, J.~A., Scott, S.~L., Tilanus, R.~P.~J., \& Wang, Z.\ 1993, \pasp, 105, 1482
\bibitem[Sch{\"o}nke \& Tscharnuter(2011)]{2011A&A...526A.139S} Sch{\"o}nke, J., \& Tscharnuter, W.~M.\ 2011, \aap, 526, A139
\bibitem[Semenov et al.(2003)]{2003A&A...410..611S} Semenov, D., Henning, T., Helling, C., Ilgner, M., \& Sedlmayr, E.\ 2003, \aap, 410, 611
\bibitem[Shu et al.(1990)]{1990ApJ...358..495S} Shu, F.~H., Tremaine, S., Adams, F.~C., \& Ruden, S.~P.\ 1990, \apj, 358, 495
\bibitem[Takahashi et al.(2006)]{2006ApJ...651..933T} Takahashi, S., Saito, M., Takakuwa, S., \& Kawabe, R.\ 2006, \apj, 651, 933
\bibitem[Takahashi et al.(2008)]{2008ApJ...688..344T} Takahashi, S., Saito, M., Ohashi, N., Kusakabe, N., Takakuwa, S., Shimajiri, Y., Tamura, M., \& Kawabe, R.\ 2008, \apj, 688, 344 (Takahashi et al. 2008a)
\bibitem[Takahashi et al.(2008)]{2008Ap&SS.313..165T} Takahashi, S., Saito, M., Takakuwa, S., \& Kawabe, R.\ 2008, \apss, 313, 165 (Takahashi et al. 2008b)
\bibitem[Takahashi et al.(2009)]{2009ApJ...704.1459T} Takahashi, S., Ho, P.~T.~P., Tang, Y.-W., Kawabe, R., \& Saito, M.\ 2009, \apj, 704, 1459
\bibitem[Takahashi \& Ho(2012)]{2012ApJ...745L..10T} Takahashi, S., \& Ho, P.~T.~P.\ 2012, \apjl, 745, L10
\bibitem[Takahashi et al.(2012)]{} Takahashi, S., Ho, P.~T.~P., Teixeira, P. S., \& Zapata, L. A., \ 2012, in preparation
\bibitem[Tomisaka(1998)]{1998ApJ...502L.163T} Tomisaka, K.\ 1998, \apjl, 502, L163
\bibitem[Tomisaka(2002)]{2002ApJ...575..306T} Tomisaka, K.\ 2002, \apj, 575, 306
\bibitem[Tomida et al.(2010)]{2010ApJ...714L..58T} Tomida, K., Tomisaka, K., Matsumoto, T., Ohsuga, K., Machida, M.~N., \& Saigo, K.\ 2010, \apjl, 714, L58 (Tomida et al. 2010a)
\bibitem[Tomida et al.(2010)]{2010ApJ...725L.239T} Tomida, K., Machida, M.~N., Saigo, K., Tomisaka, K., \& Matsumoto, T.\ 2010, \apjl, 725, L239 (Tomida et al. 2010b)
\bibitem[Zapata et al.(2010)]{2010MNRAS.402.2221Z} Zapata, L.~A., Schilke, P., \& Ho, P.~T.~P.\ 2010, \mnras, 402, 2221
\end{thebibliography}
\end{document}